Title: A Novel Mathematical/Numerical Formula for Assessing Right Ventricular Torsion Using Echocardiographic Imaging

Author: Saeed Ranjbar [1,*] Ph.D

Research institute:

1- Institute of Cardiovascular Research, Modarres Hospital, Shahid Beheshti University of Medical Sciences, Tehran, IR Iran

*Corresponding author:

Saeed Ranjbar

Modarres Hospital, Institute of Cardiovascular Research, Shahid Beheshti University of Medical Sciences, Tehran, Iran.
E-mail: sranjbar@ipm.ir Tel/FAX: +9821 22083106

# Abstract

**Background:** Recently, the ventricular torsional parameters have received special attention because of their significant role in the ventricular systolic and diastolic function. Right ventricular (RV) rotational deformation is a sensitive index for RV performance but difficult to measure. Having assumed RV as a conic shape, the present study serves a novel formula of right ventricular rotation that uses velocity vector imaging (VVI) for quantifying RV.

**Method:** After a clinical standard echocardiographic examination, the RV cross section was made as hemi ellipse as possible. After storing optimal 2D images from basal and apical RV short axis views, offline VVI analysis was performed using X-Strain software. We calculated right ventricular rotation by integrating the rotational velocity, determined from VVI global velocities of the septal and free wall regions of the RV, and correcting for the RV diameters $a(t)$ and $b(t)$ as hemi ellipse over time. Data used included: global velocities and rotational velocities of referred regions 1, 2, 3 (as septum) and 4 (as RV free wall) respectively. "a" sub 0 and "b" sub 0 are end diastolic diameters of ellipse.

**Result:** The numerical calculations showed that RV rotated with basal and apical RV rotation of $+0.75 \pm 1.2$ degree and $-2.2 \pm 2.7$ degree, respectively ($p<0.001$), and resulting twist rate of $2.95 \pm 3.9$ degree.

**Conclusion:** The present study has shown that VVI can quantify RV rotational deformation over time. This novel mathematical/numerical method may facilitate noninvasive quantification of RV twist / torsion in clinical and research settings.

# Introduction:

Previously, the right ventricle (RV) was called "the forgotten chamber", but since the recognition of the importance of the RV function in the clinical outcomes of such a wide range of conditions as pulmonary hypertension and congenital heart disease, special significance has been granted to the evaluation of its function (1). The complex geometry of the RV, however, renders the assessment of its size and function extremely challenging. Indeed, studying the relation between the RV structure and function is often likened to peeling an onion. New methods and models may have enabled us to explore the complex organization of the RV layers and fibers as well as its global chamber geometry and regional stresses and strains, but the input data for such analysis is still derived from the basic knowledge of muscle anatomy through painstaking dissection.

The RV systolic function is frequently assessed qualitatively. Although several parameters have been used for the quantitative assessment of the RV, each parameter still presents difficulties that need to be overcome (2-7). Various imaging modalities are utilized to study and characterize this anatomically complex cardiac chamber. Cardiac magnetic resonance imaging has become the reference method for the quantitative assessment of the RV size and systolic function. Nonetheless, it is an expensive method and many centers still draw on echocardiography for the RV systolic function assessment. Relatively newer methods such as strain and strain rate values and two-dimensional (2D) speckle tracking have been studied in a number of conditions. The torsion/twist of the left ventricle, defined as the wringing motion of the heart around its long axis, has been suggested as a sensitive marker of the cardiac performance. Recent studies have demonstrated the value of 2D speckle tracking and velocity vector imaging (VVI) as an advanced echocardiographic method in the evaluation of myocardial rotation and twist, but there is a paucity of information on the RV torsional parameters in the existing literature (7-13).

The purpose of the present study was to introduce a novel formula for the estimation of the basal and apical RV rotation using VVI and modeling the RV as a hemi ellipsoid.

## Method:

### RV geometrical shape assumption:

The RV fiber anatomy can be geometrically modeled as a conic when viewed from the side and a crescent in the cross-sectional views. Therefore, the RV cross-sectional images were obtained by an expert echocardiologist to make the RV as hemi-elliptically as possible.

Right ventricular adaptation cone model with RV anatomy:

RV cone assumed shape compared with MRI scans:

RV cone assumed shape compared with echocardiographic imaging (Fig. 1,2,3 and 4):

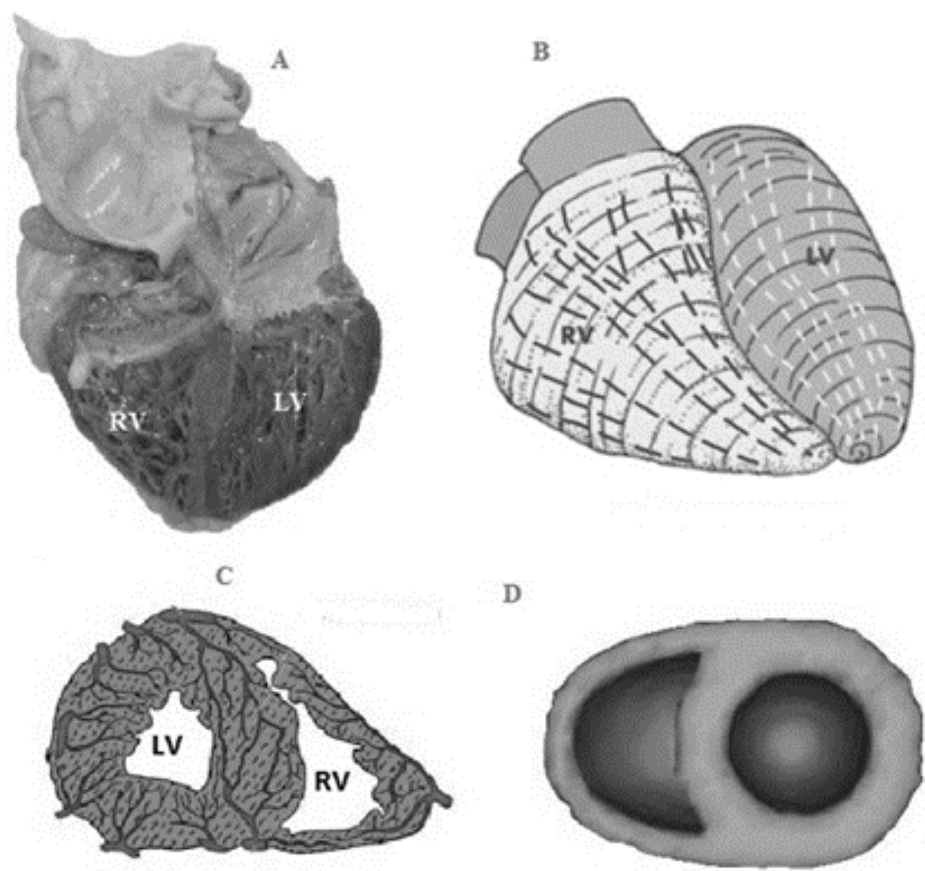

**Figure 1:** A) Open cut macroscopic view of heart in four-chamber view depicts thin-walled right ventricle, compared with left one and its heavy trabeculation. B) This panel depicts in simplistic fashion the subepicardial myofibres in the ventricles of the normal heart and this shows RV as a conic shape. ( C and D) Short axis views at the level of papillary muscles and hemi ellipse shapes of RV cross sections.

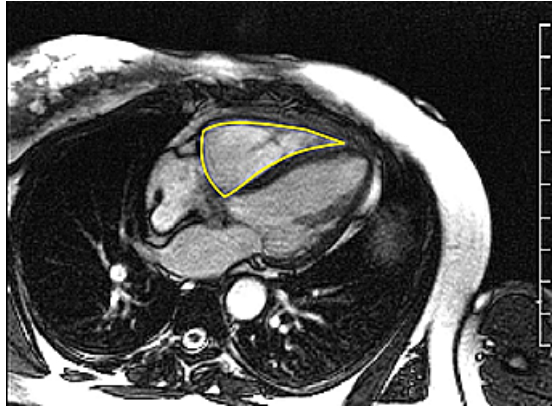

**Figure 2:** End-diastolic HLA MRI scans demonstrate the 1 major right ventricular shape : conic shaped RVs.

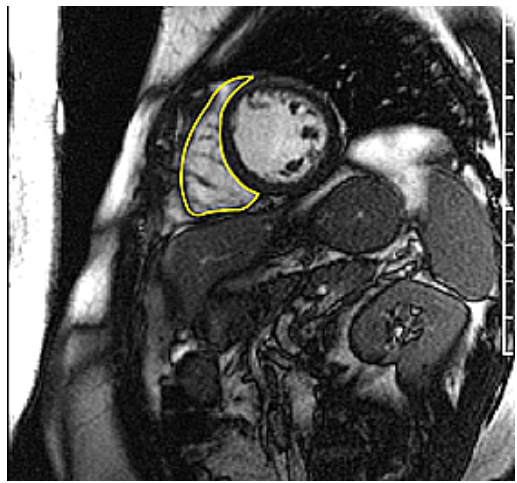

**Figure 3:** RV in short axis view at the level of papillary muscles as hemi ellipses in MR image.

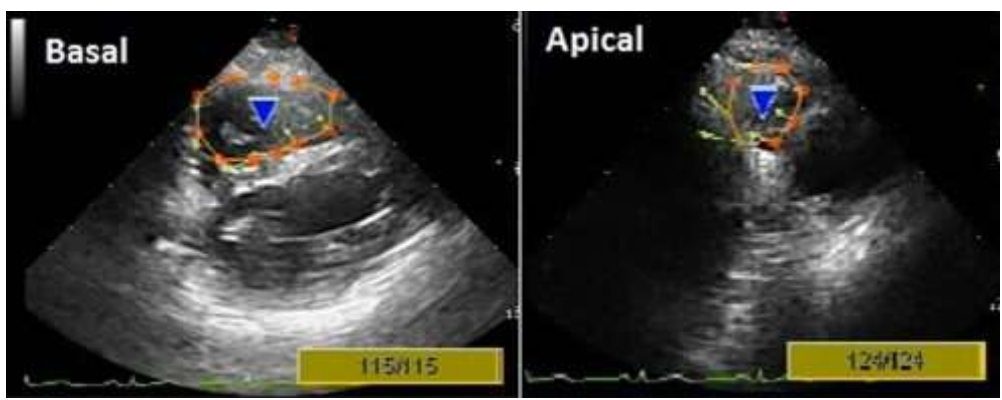

**Figure 4:** Vector velocity imaging has been applied on basal and apical short axis views: Endocardial borders with arrows showing the local directions of border motions.

**Right ventricular rotation:**

Having assumed the RV as a conic shape, we provided a novel formula for the RV rotation that uses VVI for the RV quantification.

After a clinical standard echocardiographic examination, the RV cross-section was made as elliptically as possible. Optimal 2D images from the parasternal short-axis view at the levels of the based, papillary muscles and apical for VVI analysis were recorded via electrocardiogram synchronization. We verify that we had IRB protocol approval for the use of the patient images. After storing optimal 2D images, VVI offline analysis was performed using X-Strain software (Figure 4). The RV rotation was calculated by integrating the rotational velocity, determined via VVI. The global velocities (the resultant of velocity components) of the septal and free wall regions of the RV were corrected for the RV diameters, a(t) and b(t), as an ellipse over time (Figure 5). The data used included the global velocities and rotational velocities of regions 1, 2, 3 (the septum), and 4 (the RV free wall) (Figure 5), respectively (10-18).

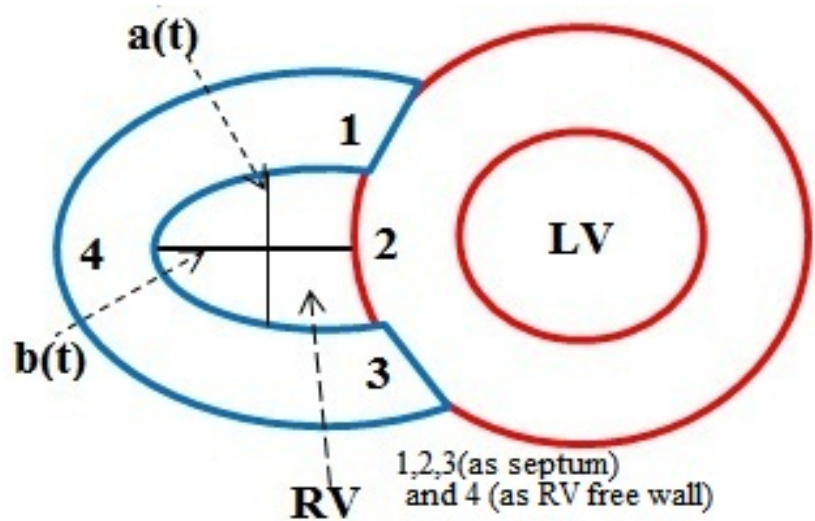

**Figure 5:** (A schematic figure of short axis view) For each short axis views at different levels (Base and apical RV), we divided RV crossed section that is as a hemi ellipse to four regions 1,2,3(as septum) and 4 (as RV free wall) which are used for RV rotational calculation in MATLAB software.

## Result:

Right ventricular rotational velocity ($RV_{rot-v}$, Rad/s) and rotation by echocardiographic velocity data sets:

$RV_{rot-v}(t)$ (degree/s) is estimated in each basal and apical RV from four points of tissue velocity data on the septum and RV free wall: 1,2,3 and 4 regions (Figure 5) to measure global velocity and rotational velocity. $RV_{rot-v}(t)$ is estimated from averaged global velocity corrected with r (t), as follows:

Eq. 1
$$RV_{rot-v}(t) = \frac{[v_4(t) - v_2(t)]}{2r(t)} \left(a(k) \cdot b(k)\right)^2$$

Where $a(k)$ and $b(k)$ are

Eq. 2
$$a(k) = \int_0^k (v(t)_{rot,1} - v(t)_{rot,3}) dt / 2$$

Eq. 3
$$b(k) = \int_0^k (v(t)_{rot,4} - v(t)_{rot,2}) dt / 2$$

And $r(t)$ is

Eq. 4
$$r(t) = \frac{a_0}{2} + \frac{b_0}{2} + \frac{\left[\int_0^{\sqrt{b(k)^2 + (1-b(k)^2)\cos^2(t)}} \int_{b(k)=0}^{b(k)=a(k)\pi} (v(t)_3 - v(t)_1)(a(k) \cdot b(k)) \left[\left(\frac{b(k)}{a(k)}\right)'\right] dtdk\right]}{2}$$

Where $v(t)_4$, $v(t)_2$, $v(t)_1$ and $v(t)_3$ are global myocardial velocities at (4),(2), (1) and (3) regions, and $v(t)_{rot,1}$, $v(t)_{rot,3}$, $v(t)_{rot,4}$ and $v(t)_{rot,3}$ are rotational velocities at (4), (2), (1) and (3) regions. $a_0$ and $b_0$ are end diastolic diameters. RV rotation (degrees) was calculated as follows:

Eq. 5
$$RV_{Rotation} = \int_0^{t_s} RV_{rot-v}(t) dt$$

Where the integration is carried from end diastole to end systole ($t_s$).

The velocity data sets of the four regions detected by Velocity vector imaging (VVI) at the basal and apical levels were then transferred to a spreadsheet program for averaging and calculation of $\text{RV}_{\text{rot-v}}(t)$. At least three consecutive cardiac cycles were averaged for these calculations (19-25).

Conclusion:

Right ventricular rotation is a critical aspect of cardiac biomechanics, although it has been difficult to measure. Our novel formula enables us to assess the RV twisty deformation using VVI. As the VVI data is obtained from relatively high-resolution two-dimensional images, the twisty velocity profile by the VVI is strongly comparable with those by the DTI method, which was derived from primary detected velocity data with higher temporal resolution but with intrinsic directionality constraints common to all Doppler techniques.

This novel method may promote noninvasive serial evaluations of the RV rotational behavior in clinical settings and provide unique and valuable information to patients with heart disease.

Reference:

1- Rudski LG, Lai WW, Afilalo J, Hua L, Handschumacher MD, Chandrasekaran K, Solomon SD, Louie EK, Schiller NB. Guidelines for the Echocardiographic Assessment of the Right Heart in Adults: A Report from the American Society of Echocardiography Endorsed by the European Association of Echocardiography, a registered branch of the European Society of Cardiology, and the Canadian Society of Echocardiography. J Am Soc Echocardiogr 2010; 23:685-713.

2- Kukulski T, Hubbert L, Arnold M, Wranne B, Hatle L, Sutherland GR. Normal regional right ventricular function and its change with age: a Doppler myocardial imaging study. J Am Soc Echocardiogr 2000;13:194 –204.

3- Eyskens B, Weidemann F, Kowalski M, Bogaert J, Dymarkowski S, Bijnens B, Gewillig M, Sutherland G, Mertens L. Regional right and left ventricular function after the Senning operation: an ultrasonic study of strain rate and strain. Cardiol Young 2004; 14:255– 64.

4- Derrick GP, Josen M, Vogel M, Henein MY, Shinebourne EA, Redington AN. Abnormalities of right ventricular long axis function after atrial repair of transposition of the great arteries. Heart 2001;86:203–6.

5- Urheim S, Edvardsen T, Torp H, Angelsen B, Smiseth OA. Myocardial strain by Doppler echocardiography. Validation of a new method to quantify regional myocardial function. Circulation 2000; 102:1158–64.

6- Fogel MA, Weinberg PM, Fellows KE, Hoffman EA. A study in ventricular-ventricular interaction. Single right ventricles compared with systemic right ventricles in a dual-chamber circulation. Circulation 1995;92:219 –30.

7- Garot J, Bluemke DA, Osman NF, Rochitte CE, McVeigh ER, Zerhouni EA, Prince JL, Lima JA. Fast determination of regional myocardial strain fields from tagged cardiac images using harmonic phase MRI. Circulation 2000; 101:981– 8.

8- Helle-Valle T, Crosby J, Edvardsen T, Lyseggen E, Amundsen BH, Smith HJ, Rosen BD, Lima JA, Torp H, Ihlen H, Smiseth OA. New noninvasive method for assessment of left ventricular rotation: speckle tracking echocardiography. Circulation. 2005;112(20):3149-56.

9- Sylvia SM Chen, Jennifer Keegan, Andrew W Dowsey, Tevfik Ismail, Ricardo Wage, Wei Li, Guang-Zhong Yang, David N Firmin, Philip J Kilner. Cardiovascular magnetic resonance tagging of the right ventricular free wall for the assessment of long axis myocardial function in congenital heart disease. Journal of Cardiovascular Magnetic Resonance. 2011;13: 2-9.

10- Ho S Y, Nihoyannopoulos P. Anatomy, echocardiography, and normal right ventricular dimensions. Heart 92 :i2–i13, 2006.

11- Fritz Jan, Solaiyappan Meiyappa, BEngg. Right Ventricle Shape and Contraction Patterns and Relation to Magnetic Resonance Imaging Findings. J Comput Assist Tomogr 2005; 29:725–733.

12- Tezuka F, Hort W, Lange PE, Nurnberg JH. Muscle fiber orientation in the development and regression of right ventricular hypertrophy in pigs. Acta Pathol Jpn 1990;40:402–7.

13- Streeter DD., Spotnitz HM, Patel DP, Ross J., Sonnenblick EH. Fiber orientation in the canine left ventricle during diastole and systole. Circ Res 1969; 24:339–47.

14- Greenbaum RA, Ho SY, Gibson DG, Becker AE, Anderson RH. Left ventricular fibre architecture in man. Br Heart J 1981;45:248–63.

15- Haber I, Metaxas DN, Geva T, Axel L. Three-dimensional systolic kinematics of the right ventricle. Am J Physiol Heart Circ Physiol 2005;289:H1826–33.

16- Plumhans C, Mühlenbruch G, Rapaee A, Sim KH, Seyfarth T, Günther RW, Mahnken AH. Assessment of global right ventricular function on 64-MDCT compared with MRI. AJR 2008; 190:1358 –1361.

17- Dupont MV, Drăgean CA, Coche EE. Right ventricle function assessment by MDCT. AJR Am J Roentgenol 2011;196(1):77–86.

18- Schepis T, Gaemperli O, Koepfli P, Valenta I, Strobel K, Brunner A, Leschka S, Desbiolles L, Husmann L, Alkadhi H, Kaufmann PA. Comparison of 64-slice CT with gated SPECT for evaluation of left ventricular function. J Nucl Med 2006;47(8):1288–1294.

19- Narayan Aravind. Perimeter of the Elliptical Arc a Geometric Method.IOSR-JM.2012;3:08-13.

20- Bruners P, Mahnken AH, Knackstedt C, Luhmann N, Spüntrup E, Das M, Hohl C, Wildberger JE, Schmitz-Rode T, Günther RW, Mühlenbruch G. Assessment of global left and right ventricular function using dual-source computed tomography (DSCT) in comparison to MRI: an experimental study in a porcine model. Invest Radiol 2007;42(11):756–764.

21- Wu YW, Tadamura E, Yamamuro M, Kanao S, Okayama S, Ozasa N, Toma M, Kimura T, Komeda M, Togashi K. Estimation of global and regional cardiac function using 64-slice computed tomography: a comparison study with echocardiography, gated-SPECT and cardiovascular magnetic resonance. Int J Cardiol 2008;128(1):69–76.

22- Marwick TH, Yu CM, Sun JP. Technical principles of tissue velocity and strain imaging methods. In: Heimdal A, editor. Myocardial imaging tissue Doppler and speckle tracking. 1st ed. Massachusette: Wiley-Blackwell; 2007.

23- Notomi Y, Lysyansky P, Setser RM, Shiota T, Popović ZB, Martin-Miklovic MG, Weaver JA, Oryszak SJ, Greenberg NL, White RD, Thomas JD. Measurement of ventricular torsion by two-dimensional ultrasound speckle tracking imaging. J Am Coll Cardiol. 2005;45(12):2034-41.

24- Idith Haber, Dimitris N. Metaxas, Tal Geva, Leon Axel. Three-dimensional systolic kinematics of the right ventricle. Am J Physiol Heart Circ Physiol 289: H1826–H1833, 2005.

25- Ingels NB, Jr., Hansen DE, Daughters GT, 2nd, Stinson EB, Alderman EL, Miller DC. Relation between longitudinal, circumferential, and oblique shortening and torsional deformation in the left ventricle of the transplanted human heart. Circ Res. 1989;64(5):915-27.